\documentclass[twocolumn,superscriptaddress,amsmath,amssymb,aps,pra,floatfix]{revtex4-2}
\usepackage[utf8]{inputenc}
\usepackage[english]{babel}
\usepackage{wrapfig}
\usepackage{float}
\usepackage{amssymb}
\usepackage{amsfonts}
\usepackage{amsmath}
\usepackage{cases}
\usepackage{stmaryrd}
\usepackage{tipa}
\usepackage[dvips]{graphicx}
\usepackage{dcolumn}
\usepackage{graphicx}
\usepackage{graphicx}
\usepackage{subfigure}
\usepackage{dcolumn}
\usepackage{bm}
\usepackage{color}
\usepackage{xcolor}
\usepackage{CJK}
\usepackage{subfigure}
\usepackage{physics}
\usepackage{array}
\usepackage{multirow}
\usepackage{mathtools}
\usepackage{enumerate}
\usepackage{comment}

\usepackage{parskip}
\usepackage{indentfirst}
\usepackage[T1]{fontenc}
\usepackage{verbatim}
\usepackage{xcolor}
\usepackage{pgf,tikz}
\usepackage{mathrsfs}

\usepackage{natbib}

\usepackage{amsmath,amssymb,textcomp}
\everymath{\displaystyle}

\usepackage{times}
\usepackage{tgheros}

\usepackage{eurosym}
\usepackage{siunitx}

\usepackage{graphicx}
\graphicspath{{./img/}}
\usepackage{svg}
\usepackage[section]{placeins}

\usepackage{hyperref}
\usepackage[normalem]{ulem}

\begin{document}

	\author{F.\,Fesquet}
	\email[]{florian.fesquet@wmi.badw.de}
	\affiliation{Walther-Mei{\ss}ner-Institut, Bayerische Akademie der Wissenschaften, 85748 Garching, Germany}
	\affiliation{School of Natural Sciences, Technische Universität München, 85748 Garching, Germany}
	
	\author{F.\,Kronowetter}
	\affiliation{Walther-Mei{\ss}ner-Institut, Bayerische Akademie der Wissenschaften, 85748 Garching, Germany}
	\affiliation{School of Natural Sciences, Technische Universität München, 85748 Garching, Germany}
	\affiliation{Rohde \& Schwarz GmbH \& Co. KG, 81671 Munich, Germany}
	
	\author{M.\,Renger}
	\affiliation{Walther-Mei{\ss}ner-Institut, Bayerische Akademie der Wissenschaften, 85748 Garching, Germany}
	\affiliation{School of Natural Sciences, Technische Universität München, 85748 Garching, Germany}
	
		\author{W.\,K.\,Yam}
	\affiliation{Walther-Mei{\ss}ner-Institut, Bayerische Akademie der Wissenschaften, 85748 Garching, Germany}
	\affiliation{School of Natural Sciences, Technische Universität München, 85748 Garching, Germany}
	
		\author{S.\,Gandorfer}
	\affiliation{Walther-Mei{\ss}ner-Institut, Bayerische Akademie der Wissenschaften, 85748 Garching, Germany}
	\affiliation{School of Natural Sciences, Technische Universität München, 85748 Garching, Germany}
	
		\author{K.\,Inomata}
	\affiliation{RIKEN Center for Quantum Computing (RQC), Wako, Saitama 351-0198, Japan}
	\affiliation{National Institute of Advanced Industrial Science and Technology,
		1-1-1 Umezono, Tsukuba, Ibaraki, 305-8563, Japan}
	
		\author{Y.\,Nakamura}
	\affiliation{RIKEN Center for Quantum Computing (RQC), Wako, Saitama 351-0198, Japan}
	\affiliation{Department of Applied Physics, Graduate School of Engineering,
		The University of Tokyo, Bunkyo-ku, Tokyo 113-8656, Japan}
	
	\author{A.\,Marx}
	\affiliation{Walther-Mei{\ss}ner-Institut, Bayerische Akademie der Wissenschaften, 85748 Garching, Germany}	
	
	\author{R.\,Gross}
	\affiliation{Walther-Mei{\ss}ner-Institut, Bayerische Akademie der Wissenschaften, 85748 Garching, Germany}
	\affiliation{School of Natural Sciences, Technische Universität München, 85748 Garching, Germany}
	\affiliation{Munich Center for Quantum Science and Technology (MCQST), 80799 Munich, Germany}
	
	\author{K.\,G.\,Fedorov}
	\email[]{kirill.fedorov@wmi.badw.de}
	\affiliation{Walther-Mei{\ss}ner-Institut, Bayerische Akademie der Wissenschaften, 85748 Garching, Germany}
	\affiliation{School of Natural Sciences, Technische Universität München, 85748 Garching, Germany}
	\affiliation{Munich Center for Quantum Science and Technology (MCQST), 80799 Munich, Germany}
	
	\title{Demonstration of microwave single-shot quantum key distribution}

\begin{abstract}
	Security of modern classical data encryption often relies on computationally hard problems, which can be trivialized with the advent of quantum computers. A potential remedy for this is quantum communication which takes advantage of the laws of quantum physics to provide secure exchange of information. Here, quantum key distribution (QKD) represents a powerful tool, allowing for unconditionally secure quantum communication between remote parties. At the same time, microwave quantum communication is set to play an important role in future quantum networks because of its natural frequency compatibility with superconducting quantum processors and modern near-distance communication standards. To this end, we present an experimental realization of a continuous-variable QKD protocol based on propagating displaced squeezed microwave states. We use superconducting parametric devices for generation and single-shot quadrature detection of these states. We demonstrate unconditional security in our experimental microwave QKD setting. We show that security performance can be improved by adding finite trusted noise to the preparation side. Our results indicate feasibility of secure microwave quantum communication with the currently available technology in both open-air (up to $\sim \SI{80}{\metre}$) and cryogenic (over \SI{1000}{\metre}) conditions.
\end{abstract}
	
	\maketitle

\noindent
Quantum key distribution (QKD) is a method to securely exchange information between two authenticated remote parties. Contrary to classical encryption relying on computationally asymmetric tasks, security of QKD protocols is based on quantum mechanical properties. Among the variety of existing QKD protocols, continuous-variable (CV) protocols have been extensively developed due to their technological compatibility with existing classical communication platforms, their ability to deliver high secret key rates over large distances, and less demanding experimental requirements as compared to discrete-variable protocols~\cite{Xu2020,Pirandola2015,Pirandola2020}. In the optical domain, CV-QKD protocols have been successfully implemented within large networks and achieved high secure bit rates~\cite{Peev2009,Liao2017,Chen2021,Wang2022}. In parallel, a tremendous progress has been made in quantum information processing with superconducting circuits operating at microwave frequencies~\cite{Arute2019, Bienfait2019,Pogorzalek2019, Kjaergaard2020,Fedorov21, Kronowetter2023}. Arguably, this field holds the biggest promise to achieve scalable quantum computing. Therefore, microwave CV-QKD protocols possess a huge potential due to their intrinsic frequency and technology compatibility with superconducting quantum processors, while providing access to unconditionally secure communication. Recent theoretical studies~\cite{Pirandola2021,Fesquet2023} indicate that microwave CV-QKD protocols can be implemented in open-air conditions, potentially complementing short-distance classical communication protocols such as WiFi, Bluetooth, or even 5G. There, microwave communication benefits from a strong resilience to weather conditions, as compared to optical communication~\cite{Zhao2000, Kaushal2018}.
	
\begin{figure*}[ht]
\centering
\includegraphics[width=1\textwidth]{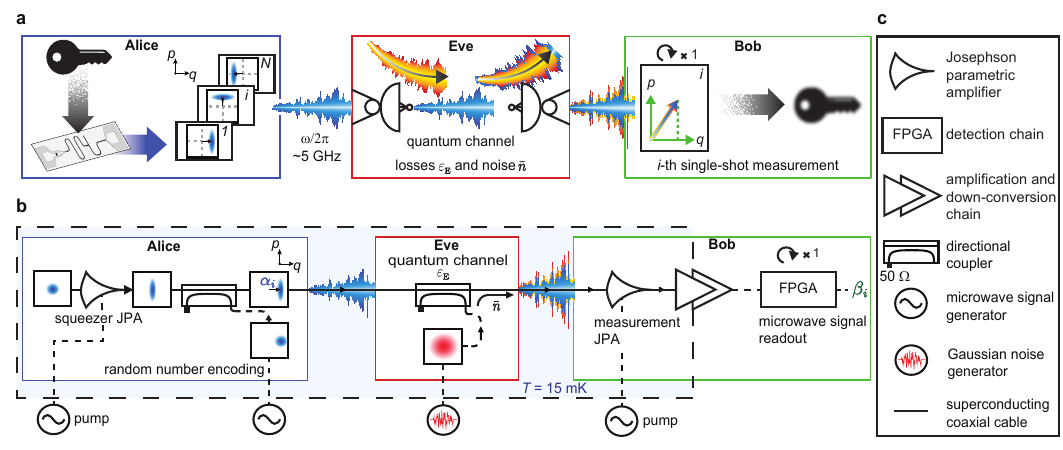}
\caption{General concept of a prepare-and-measure CV-QKD protocol based on displaced squeezed states and its microwave experimental implementation. $\textbf{a}$, In the CV-QKD protocol, Alice encodes her key $\mathcal{K}_{\mathrm{A}} = \left\lbrace \alpha_{i}\right\rbrace_{i \in \left\lbrace 1,\ldots, N\right\rbrace}$ in an ensemble of \textit{q}- or \textit{p}-displaced squeezed states. These states propagate as microwave signals through a quantum channel, which is assumed to be under Eve's control and is parametrized by power losses $\varepsilon_{\mathrm{E}}$ and added noise photon number $\bar{n}$. Bob performs SQMs to extract displacement amplitudes of each incoming state, resulting in a measured key $\mathcal{K}_{\mathrm{B}} = \left\lbrace \beta_{i}\right\rbrace_{i \in \left\lbrace 1,\ldots, N\right\rbrace}$. $\textbf{b}$, Experimental scheme of the microwave CV-QKD protocol with superconducting JPAs in the cryogenic environment. For each symbol, Alice generates a $q$- or $p$-squeezed state which is subsequently displaced using a directional coupler coupled to a strong coherent signal. The resulting state propagates through a quantum channel consisting of a second directional coupler with transmissivity $1-\varepsilon_{\mathrm{E}} = 0.9885$. This coupler is used to inject a variable number of noise photons $\bar{n}$ and, thus, simulate different channel conditions. On Bob's side, a strong phase-sensitive amplification is performed using a second JPA, resulting in the SQM of each received microwave signal. Each of these signals is sampled using a field-programmable gate array (FPGA) to compute a single \textit{I/Q} point from which the displacement $\beta_{i}$ is obtained. Color plots in boxes depict Wigner functions of quantum states in the quadrature phase space $(q,p)$. $\textbf{c}$,~Legend for various experimental components in panel $\textbf{b}$.}
\label{fig:Fig_1}
\end{figure*}
\par A general CV-QKD protocol aims to securely exchange information between a sender (Alice) and a receiver (Bob) using coherent or squeezed states. Information is encoded as a sequence of numbers, referred to as a key, in the $q$- and $p$-field quadrature bases of these states. The quantum states propagate through a quantum channel, which is assumed to be under the full control of a malicious eavesdropper (Eve) who tries to syphon information about the key. The security of CV-QKD protocols relies on a single use of each state prepared by Alice, since weak averaging measurements of multiple copies reveal too much information to Eve and compromise the unconditional security~\cite{Pirandola2008_2}. For protocols based on squeezed states, where information is encoded into a single field quadrature, Bob implements single-shot quadrature measurements (SQMs) of the encoding quadrature. In the optical domain, this task is conventionally performed using a homodyne detection technique. In the microwave domain, we achieve an equivalent signal detection using superconducting phase-sensitive amplifiers~\cite{Yurke1989,Castellanos-Beltran2008,Eichler2012,Mallet2011}. After successfully extracting the key from SQMs, security of the communication must be assessed to determine whether Alice and Bob can obtain a secure shared key.

In this work, we present an experimental realization of a one-way CV-QKD protocol based on the Gaussian modulation of propagating squeezed microwave states~\cite{Cerf2001} in a cryogenic environment. Our experiment serves as a proof of principle for microwave CV-QKD protocols and sheds light on their practical limitations. For SQMs, we use a superconducting Josephson parametric amplifier (JPA), which enables strong phase-sensitive amplification and high quantum efficiency well beyond the standard quantum limit~\cite{Yamamoto2008,Zhong2013,Renger2021}. We focus on a trusted-device scenario, where preparation losses and a detection noise are trusted. The quantum channel is an untrusted lossy and noisy channel, experimentally implemented by a cryogenic directional coupler (highly asymmetric microwave beam splitter) with fixed power losses $\varepsilon_{\mathrm{E}}$ and tunable coupled Gaussian noise. The latter is characterized by $\bar{n}$ artificially generated noise photons coupled to the propagating signal. As such, our experiment can be viewed as a quantum simulation of a real CV-QKD implementation, where we controllably vary the temperature of the thermal background in the quantum channel. To prove security of the protocol, we study the worst-case scenario, where Eve exploits a collective Gaussian attack over the ensemble of states sent by Alice~\cite{Garcia-Patron2006}. Our analysis demonstrates a feasibility of unconditionally secure microwave CV-QKD in a cryogenic environment over distances approaching $\SI{1200}{\metre}$, which corresponds to open-air conditions with secure communication distances up to $\SI{80}{\metre}$. Owing to the finite length of the exchanged keys, we extend our security analysis by considering both conventional finite-size induced terms~\cite{Leverrier2010} and quantum channel parameter estimations~\cite{Laudenbach2018}. Here, we experimentally demonstrate secure communication for a key length of $N = 16665$ numbers, also commonly referred to as symbols. We find that our experiment allows for an accurate statistical estimation of the channel losses and coupled noise.
 
\section*{Microwave CV-QKD protocol implementation}
\noindent
Our CV-QKD protocol relies on the generation of displaced squeezed microwave states to encode a key from Alice. In Fig.~\ref{fig:Fig_1}, we illustrate its concept and present a microwave scheme of our experimental implementation. Here, we choose the carrier frequency of all quantum states to be $\omega/2\pi = \SI{5.48}{\giga\hertz}$. We use a superconducting flux-driven JPA for generation of squeezed microwave states, which are characterized by a squeezing level $S$ below vacuum~\cite{Fedorov2016,Menzel2012}. Our JPAs consist of a coplanar waveguide $\lambda /4$ resonator short-circuited to ground by a direct current superconducting quantum interference device (dc-SQUID). The dc-SQUID provides a flux-tunable inductance, which allows for frequency tuning of the JPAs. This flux tunability is the key for parametric amplification of microwave signals and generation of squeezed states. For the latter, our JPAs are operated in the phase-sensitive regime by pumping them at twice their resonance frequencies $\omega_{\mathrm{p}} = 2\omega$. The squeezed states are subsequently displaced in quadrature phase space using a cryogenic directional coupler~\cite{Fedorov2016}. Each displacement operation encodes a symbol $\alpha_i$ drawn from a codebook following Gaussian distribution with the fixed variance $\sigma_{\mathrm{A}}^2$. These symbols constitute  Alice's key $\mathcal{K}_{\mathrm{A}} = \left\lbrace \alpha_{i}\right\rbrace_{i \in \left\lbrace 1,\ldots, N\right\rbrace}$. Displacement and squeezing operations are performed either along the $q$ or $p$ directions in phase space, chosen randomly for each symbol. For maximal security, the codebook variance is calibrated such that averaging over the ensemble of Alice's states results in a thermal state, preventing Eve from extracting information on the encoding basis. This imposes the condition $\sigma_{\mathrm{s}}^2 + \sigma_{\mathrm{A}}^2 = \sigma_{\mathrm{as}}^2$, where $\sigma_{\mathrm{s}}^2$ ($\sigma_{\mathrm{as}}^2$) denotes the squeezed (anti-squeezed) quadrature variances. In our measurements, we keep a constant squeezing level $S = \SI{3.6\pm 0.4}{\decibel}$. For signal readout, Bob uses a second JPA to perform the SQMs with a quantum efficiency that depends on the added JPA noise. This noise is related to intrinsic losses, pump-induced noise~\cite{Renger2021,Fedorov21}, and higher-order nonlinearities~\cite{Boutin2017}. Single-shot measurements, ideally implemented with quantum efficiency close to unity, are obtained with a quantum efficiency well above $50\,\%$ and without any averaging of measured signals. The SQM is performed for each symbol encoded by Alice and results in a measured key for Bob $\mathcal{K}_{\mathrm{B}} = \left\lbrace \beta_{i}\right\rbrace_{i \in \left\lbrace 1,\ldots, N\right\rbrace}$. In practical implementations, a CV-QKD protocol includes additional post-processing, notably, a classical error correction algorithm which uses either Alice's or Bob's keys as a reference to provide them with a common key. Here, we consider the direct reconciliation (DR) regime, where Alice's key is used as a reference. This regime is known to offer a better resilience to the coupled noise $\bar{n}$ as compared to reverse reconciliation, where Bob's key is taken as the reference~\cite{Grosshans2003,Grosshans2003_2}. 
\section*{Single-shot measurements and correlations}
\noindent
To describe the strong phase-sensitive amplification resulting in SQMs, we use the covariance matrix formalism. When the $q$ quadrature is amplified, we write the covariance matrix of an amplified single-mode state as 
\begin{equation}
	\label{eq:amplification}
	\boldsymbol{V}^{'} = \boldsymbol{J}^{\mathrm{T}}\, \boldsymbol{V}\, \boldsymbol{J} + \boldsymbol{N} \text{,} \quad \boldsymbol{J} = \begin{pmatrix}
		\sqrt{G_{\mathrm{J}}} & 0 \\
		0 & 1/\sqrt{G_{\mathrm{J}}}\\
	\end{pmatrix} \text{,} 
\end{equation}
where $G_{\mathrm{J}}$ is the degenerate JPA gain and $\boldsymbol{V}$ is the input covariance matrix. A similar equation can be written for amplification along the $p$ quadrature by swapping the diagonal terms of $\boldsymbol{J}$. Additionally, $\boldsymbol{N}$ is a diagonal matrix representing the noise added by our amplification chain. From Eq.\,\ref{eq:amplification}, we find that the first diagonal element, corresponding to the $q$ quadrature, is enlarged by the degenerate gain as $V'_{11} = G_{\mathrm{J}} V_{11} + N_{11}$. Conversely, the second diagonal element, corresponding to the $p$ quadrature, is attenuated by the degenerate gain to $V'_{22} = V_{22}/G_{\mathrm{J}} + N_{22}$. As a result, in the case of large gain, $G_{\mathrm{J}} \gg 1$, and finite amplification noise, information about the deamplified quadrature becomes inaccessible from single-shot quadrature measurements as opposed to the amplified quadrature.
	\begin{figure*}[htb]
	\includegraphics[width=1\textwidth]{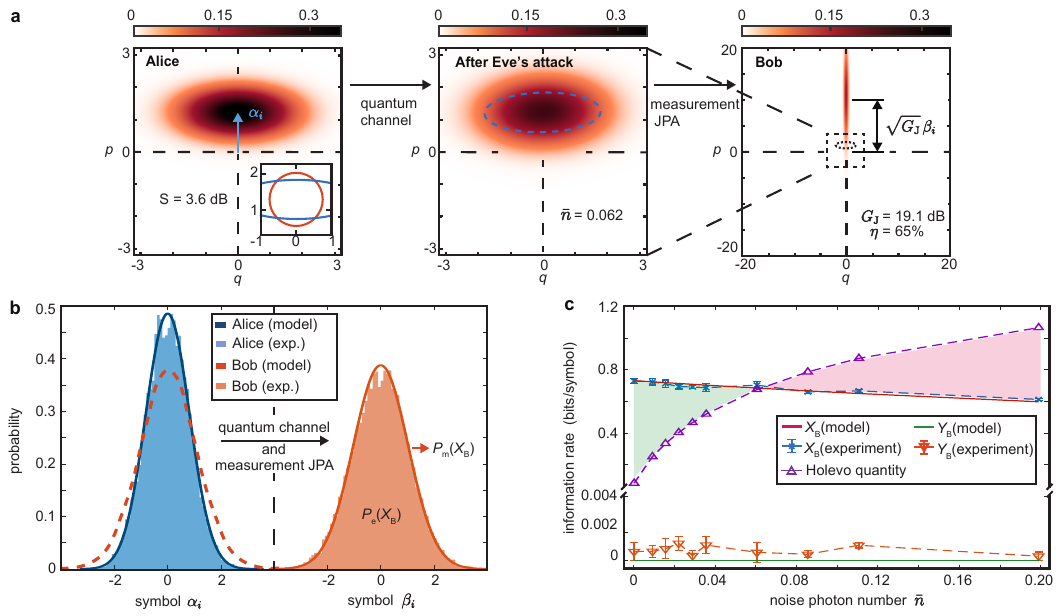}
	\caption{Tomography and single-shot measurement histograms of displaced squeezed microwave states. $\textbf{a}$, Exemplary reconstructed Wigner function of the evolution of a quantum key symbol, starting from its preparation at Alice, followed by propagation through the quantum channel while being exposed to losses and noise (Eve's attack), finishing at Bob with a strong phase-sensitive amplification. The inset of the left Wigner function plot shows the 1/e contours for an ideal vacuum (red circle) and experimental squeezed state (blue ellipsoid) indicating squeezing below the level of vacuum fluctuations. $\textbf{b}$, Exemplary measured histograms for Alice's and Bob's key symbols. For comparison with the measured probability distribution~$P_{\mathrm{e}}\left(X_{\mathrm{B}}\right)$, we plot our quadrature model (solid lines) resulting in a zero-mean Gaussian probability distribution $P_{\mathrm{m}}\left(X_{\mathrm{B}}\right)$, whose variances are obtained from independent calibration measurements (see Methods). $\textbf{c}$, MI between Alice's and Bob's keys for the amplified (deamplified) quadrature $X_{\mathrm{B}}$~ $\left(Y_{\mathrm{B}}\right)$ as a function of the coupled noise photon number $\bar{n}$. We additionally show the MI computed from our model, which is also based on the independent calibration measurements. We emphasize that the model is not a fit of the measurement data. Lastly, we show the corresponding Holevo quantity. The shaded green (red) area represents the region where the MI is larger (smaller) than the Holevo quantity, resulting in a unconditionally secure (insecure) communication.} 
	\label{fig:Fig_2}
\end{figure*}
Experimentally, we characterize the quadrature amplification noise $\bar{n}_{\mathrm{x}}$ using the quantum efficiency $\eta =1/(1+2 \bar{n}_{\mathrm{x}})$~\cite{Boutin2017}, which we optimize to be as close as possible to unity. In Fig.~\ref{fig:Fig_2}$\textbf{b}$, we show an exemplary normalized histogram of single-shot measurements of Bob's symbols with $G_{\mathrm{J}} = \SI{19.1\pm 0.4}{\deci\bel} $ and $\eta =65\pm 2~\%$. Superimposed to the histogram, we plot an extrapolated quadrature distribution model (see Methods) based on the formalism in Eq.\,\ref{eq:amplification}. The numerical values of our model parameters are obtained from independent calibration measurements (see Methods) of the experimental setup components. We show in Fig.~\ref{fig:Fig_2}$\textbf{a}$ the evolution of an exemplary Wigner state tomography of a displaced squeezed state. 

Following these measurements, Bob possesses a set of symbols correlated to the initial set sent by Alice. We characterize these correlations by computing the mutual information (MI) between Alice's encoded key $\mathcal{K}_{\mathrm{A}}$ and Bob's corresponding measured key $\mathcal{K}_{\mathrm{B}}$. For continuous-variable states, the MI, assuming SQMs, is expressed as
\begin{equation}
	\label{eq:MI}
	I\left(\mathcal{K}_{\mathrm{A}}\colon \mathcal{K}_{\mathrm{B}}\right) = h\left(\mathcal{K}_{\mathrm{B}}\right) - h\left(\mathcal{K}_{\mathrm{B}}\vert \mathcal{K}_{\mathrm{A}}\right) = \frac{1}{2}\log_2\left(1+ \mathrm{SNR} \right) \text{,}
\end{equation}
where $h$ is the differential entropy and $\mathrm{SNR}$ is the signal-to-noise ratio. In Fig.~\ref{fig:Fig_2}$\textbf{c}$, we plot the MI extracted from our measurement for the amplified (deamplified) quadrature, denoted as $X_{\mathrm{B}}$ ($Y_{\mathrm{B}}$). We note that the MI is insensitive to any linear rescaling of either Alice's or Bob's keys and, therefore, captures core correlations between their datasets. For the quadrature $X_{\mathrm{B}}$, we observe a clearly non-zero MI, indicating strong correlations between Alice's and Bob's key. Conversely, we observe a nearly zero MI for the deamplified quadrature, demonstrating the almost complete loss of information, as expected from the Heisenberg principle when measuring conjugate quantum variables.  Additionally, we show values for the MI based on our model under the assumption that Alice's and Bob's keys follow a Gaussian distribution. The accuracy of our model is quantified using the Bhattacharyya coefficient, $\mathcal{B}$, which we use to evaluate the overlap between measured quadrature distributions and our corresponding model predictions. The coefficient $\mathcal{B}$ can be viewed as the classical analogue of the Ulhmann fidelity of density matrices~\cite{fuchs1998cryptographic} and provides a well-established metric for probability distributions, the Hellinger distance $H(P_1,P_2) = \sqrt{1-\mathcal{B}(P_1,P_2)}$, where $P_1$ and $P_2$ denote two different probability distributions. The quantum counterpart of the Hellinger distance is closely related to the trace distance between density matrices~\cite{fuchs1998cryptographic}. Based on the measurements presented in Fig.\,\ref{fig:Fig_2}, we compute the coefficient $\mathcal{B}(P_{\mathrm{e}}\left(X_{\mathrm{B}}\right),P_{\mathrm{m}}\left(X_{\mathrm{B}}\right)) = \SI{99.97 \pm 0.01}{\percent}$ with an associated $H(P_{\mathrm{e}}\left(X_{\mathrm{B}}\right),P_{\mathrm{m}}\left(X_{\mathrm{B}}\right)) = \num{0.017\pm 0.003}$ for $P_{\mathrm{e}}\left(X_{\mathrm{B}}\right)$ the probability distribution of the experimentally measured amplified quadrature $X_{\mathrm{B}}$ and $P_{\mathrm{m}}\left(X_{\mathrm{B}}\right)$ its corresponding quadrature distribution predicted from our model. Replacing the quadrature $X_{\mathrm{B}}$ by the deamplified quadrature $Y_{\mathrm{B}}$ results in $\mathcal{B}(P_{\mathrm{e}}\left(Y_{\mathrm{B}}\right),P_{\mathrm{m}}\left(Y_{\mathrm{B}}\right)) = \SI{99.92 \pm 0.01}{\percent}$ with $H(P_{\mathrm{e}}\left(Y_{\mathrm{B}}\right),P_{\mathrm{m}}\left(Y_{\mathrm{B}}\right)) = \num{0.028 \pm 0.002}$. The near-zero Hellinger distances and associated $\mathcal{B}$ values close to unity indicate excellent agreement between our quadrature distribution model and experimental measurements, which can be interpreted as a proof for genuine single-shot quadrature measurements in our experiments. 
\section*{security analysis}
\begin{figure}[t]
	\centering
	\includegraphics[width=1\columnwidth]{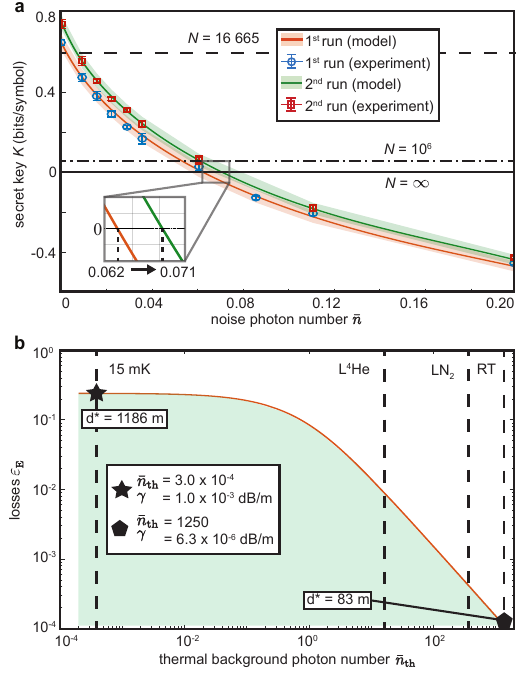}
	\caption{Secret key of the microwave CV-QKD protocol. $\textbf{a}$, Measured 
		secret key of the CV-QKD protocol for two experimental runs: $1\textsuperscript{st}$ with squeezing (anti-squeezing) levels of  3.6 (7.1) dB and $2\textsuperscript{nd}$ with squeezing (anti-squeezing) levels of  3.6 (7.6) dB. Importantly, the $2\textsuperscript{nd}$ run delivers a higher secret rate, illustrating the positive impact of increased trusted noise on Alice's side, and thus, the codebook size, on the secret key. The shaded areas denote the standard deviation of our model. The dashed lines represent the finite-size terms, which impose lowered noise cut-offs for reaching the unconditional security. $\textbf{b}$, Estimation of maximally tolerable losses (solid line) for positive secret keys as a function of the photon number in the thermal background,  $\bar{n}_{\mathrm{th}}$. This analysis is based on the experimental data from the $2\textsuperscript{nd}$ run. The green shaded area indicates the region of positive (i.e., secure) secret keys. We emphasize two particular temperatures on this curve: the cryogenic temperature $\sim \SI{15}{\milli\kelvin}$ and room temperature (RT) $\sim \SI{300}{\kelvin}$. At millikelvin temperatures, we assume characteristic losses in superconducting cables of $\gamma = \SI{1.0e-3}{\decibel /\meter}$ while for the open-air conditions, we restrict ourselves to atmospheric microwave losses $\gamma = \SI{6.3e-6}{\decibel / \meter}$ due to pure absorption.  Under these conditions, we estimate the maximum communication distance, $d^* = \SI{1186}{\metre}$ at \SI{15}{\milli\kelvin} and $d^* = \SI{86}{\metre}$ at \SI{300}{\kelvin}. For the open-air scenario, we neglect possible path losses, assuming those can be fully compensated by appropriate antennae, and focus on unavoidable physical limitations.}
	\label{fig:Fig_3}
\end{figure}

\noindent
In order to extract secret information from their datasets, Alice and Bob need to estimate an upper bound for the amount of information leaked during the quantum communication using the Holevo quantity $\chi_{\mathrm{E}}$. First, we consider the asymptotic case where communicated keys are assumed to be infinitely long. In this case, we rely on our calibration measurements to have an exact knowledge about the channel losses and coupled noise. In Fig.\,\ref{fig:Fig_2} $\textbf{c}$, we show the resulting Holevo quantity for our presented protocol implementation. Without a loss of generality~\cite{Garcia-Patron2006,Renner2009}, we can assume that Eve employs a collective Gaussian attack~\cite{Pirandola2008}, where she interacts with all states from Alice and stores them in perfect quantum memories before applying an optimal joint measurement. Notably, the Holevo quantity has the advantage of being independent of any joint measurements made by Eve. Under these considerations, we can restrict her collective Gaussian attack to an entangling cloner attack~\cite{Grosshans2003_2}, where Eve couples each incoming signal of Alice to one mode of a two-mode squeezed state. From the perspective of Alice and Bob, Eve's coupled signal appears as a thermal noise signal with $\bar{n}_{\mathrm{th}} = 2\bar{n}/\varepsilon_{\mathrm{E}}$. Knowing the maximum information leaked to Eve, Alice and Bob can assess the security of communication in the asymptotic case by bounding the number of secure bits communicated per symbol $K_{\mathrm{exp}}$ with the secret key $K= I\left(\mathcal{K}_{\mathrm{A}}\colon \mathcal{K}_{\mathrm{B}}\right) - \chi_{\mathrm{E}} \leq K_{\mathrm{exp}}$. In Fig.\,\ref{fig:Fig_3} $\textbf{a}$, we show the secret key $K$ associated with the MI presented in Fig.\,\ref{fig:Fig_2} $\textbf{c}$. We observe a clear positive secret key, which indicates that Alice and Bob share more information than what leaks to Eve.
Thus, our microwave CV-QKD protocol achieves unconditional security in the asymptotic regime. More precisely, the secret key remains positive up to $\num{0.062 \pm 0.002}$ coupled noise photons. Different approaches can be chosen to improve the protocol performance, mainly by increasing the codebook size, squeezing level, or quantum efficiency. However, various limitations, such as compression effects of the JPAs, JPA noise performance, finite losses, and experimentally achievable squeezing levels, must be taken into account. In our experiments, we can enlarge the codebook variance $\sigma_{\mathrm{A}}^2$ by allowing for additional input noise from the first JPA while simultaneously keeping the squeezing level constant. This results in an increase of the anti-squeezing level from $\SI{7.1}{\deci\bel}$ to $\SI{7.6}{\deci\bel}$ and, hence, in an enhancement of $\sigma_{\mathrm{A}}^2$ by $\sim 14 \%$. As shown in Fig.\,\ref{fig:Fig_3}\, $\textbf{a}$, this increased codebook variance leads to a higher secret key, extending the noise tolerance to $\num{0.071 \pm 0.002}$ photons. During this 2\textsuperscript{nd} run, we also obtain a slightly higher quantum efficiency of $\eta =68 \pm 2~\%$ as compared to the initial $\eta = 65 \pm 2~\%$. However, based on our quadrature distribution model, this increase in $\eta$ alone is insufficient to induce the observed higher secret keys. Therefore, our experiment demonstrates that adding a finite amount of trusted noise on Alice's side can lead to increased secure key values. This result illustrates a general beneficial effect of adding trusted noise on the reference side of error correction~\cite{Garcia-Patron2009}. For the case of the lowest coupled noise, $\bar{n} \simeq 1.7 \times 10^{-6}$ (given by the coupling to our sample stage at $T \simeq \SI{15}{\milli\kelvin}$), we measure a relatively high secret key up to $0.74$ bits/symbol and, according to Eq.\,\ref{eq:MI}, a corresponding SNR of $2.16$, similar to optical implementations in long-distance communication \cite{Jouguet2011}.

Our security analysis can be extended to include limiting effects arising from the finite size of the transmitted key~\cite{Leverrier2010}. These finite-size effects induce a decrease of the secret key and are reflected by additional finite-size terms~$\Delta$ (see Supplementary Information). Equally important, we must account for the fact that in practical QKD implementations, Alice and Bob do not have exact knowledge on the quantum channel parameters and must estimate these parameters using part of the communicated key. To achieve maximal security, the channel parameters are obtained from worst-case-scenario statistical estimators $\varepsilon_{\mathrm{E}}^{\star}$ and $\bar{n}^{\star}$ for the channel losses $\varepsilon_{\mathrm{E}}$ and coupled noise $\bar{n}$, respectively. Following the approach in ref.\,\citenum{Pirandola2021}, the secret key bound takes the form $r\left[\beta I\left(\mathcal{K}_{\mathrm{A}}\colon \mathcal{K}_{\mathrm{B}}\right) - \chi_{\mathrm{E}}\left(\varepsilon_{\mathrm{E}}^{\star}, \bar{n}^{\star}\right)  -\Delta\left(n_{\mathrm{exp}}\right) \right]\leq K_{\mathrm{exp}}$, where $r = n_{\mathrm{ec}}\,p_{\mathrm{ec}}/N$ is a rescaling prefactor with $n_{\mathrm{ec}}$ denoting the fraction of the exchanged key which is not used for parameter estimation. The efficiency of the error correction protocol is denoted as $\beta$ with its success probability $p_{\mathrm{ec}}$.  We note that for CV-QKD protocols, this efficiency can reach $\beta > 90 \%$~\cite{Jouguet2013} for an $\mathrm{SNR}$ around unity. In our protocol, we use the key length of $N = 16665$ symbols. As illustrated in Fig.~\ref{fig:Fig_3} $\textbf{a}$, if we account only for the finite-size terms $\Delta$, we can observe a region of positive secret key up to $\bar{n} = 0.005$ ($\bar{n} = 0.01$) for the 1\textsuperscript{st} run (2\textsuperscript{nd} run). These effects can be largely mitigated by extending the key length to a more demanding but realistic value of $N \geq 10^6$. From our experimental keys, we compute a worst-case unbiased estimator for the losses and noise as $\varepsilon_{\mathrm{E}}^{\star} = \hat{\varepsilon}_{\mathrm{E}} - w\sigma_{\hat{\varepsilon}_{\mathrm{E}}}$ and $\bar{n}^{\star} = \hat{\bar{n}} + w\sigma_{\hat{\bar{n}}}$, with unbiased estimators $\hat{\varepsilon}_{\mathrm{E}}$ and $\hat{\bar{n}}$, built using $N-n_{\mathrm{ec}}$ symbols of Alice's and Bob's key. Here, $w$ is a confidence parameter for a chosen statistical error $\varepsilon_{\mathrm{ec}}$ reducing to $w = \sqrt{2}\, \text{erf}^{-1}\left(1-2 e_{\mathrm{ec}} \right)$ in the case of Gaussian variables. Considering a typical CV-QKD error of $e_{\mathrm{ec}} = 10^{-10}$ and not accounting for the finite-size terms $\Delta$, we vary the fraction $N - n_{\mathrm{ec}}$ to build the estimators $\varepsilon_{\mathrm{E}}^{\star}$ and $\bar{n}^{\star}$, leading to a positive secret key up to roughly $\bar{n} = 0.02$ ($\bar{n} = 0.03$) for the 1\textsuperscript{st} run (2\textsuperscript{nd} run). We conclude that all finite-size effects can be straightforwardly solved by increasing the key length to $N \geq 10^6$ and for typical values $e_{\mathrm{ec}}$ in CV-QKD protocols.

Finally, to provide a more application-oriented outlook, we estimate maximal communication distances the microwave CV-QKD protocol could achieve with the current experimental performance. To this end, we consider a communication protocol, where Alice and Bob keep the same experimental parameters as in the $2\textsuperscript{nd}$ run, except for modified losses $\varepsilon_{\mathrm{E}}$ and noise photon numbers $\bar{n} = \bar{n}_\mathrm{th} \varepsilon_{\mathrm{E}} / 2 $ of the quantum channel. In Fig.~\ref{fig:Fig_3}$\textbf{b}$, we show maximally tolerable losses for a given photon number $\bar{n}_{\mathrm{th}}$. We find that the unconditionally secure microwave communication up to $\SI{1186}{\meter}$ is feasible in a fully cryogenic environment at $T \simeq \SI{15}{\milli\kelvin}$  based on commercial superconducting cables with characteristic losses of $\SI{1.0e-3}{\decibel /\meter}$~\cite{Kurpiers2017}, making microwave CV-QKD relevant for secure local area quantum networks~\cite{Awschalom2021}. We also find that the unconditionally secure microwave communication should be possible up to $\SI{83}{\meter}$ in the open-air environment with $\bar{n}_{\mathrm{th}} \simeq 1250$ for signals at $\omega /2\pi = \SI{5}{\giga\hertz}$. This finding results from considering the very low microwave atmospheric absorption losses of $\SI{6.3e-6}{\decibel /\meter}$~\cite{Fesquet2023} in clear weather conditions, primarily limited by oxygen and water absorption. As such, microwave CV-QKD demonstrates a notable potential for secure short-range open-air microwave communication, where microwave signals additionally benefit from a resilience to weather imperfections \cite{Fesquet2023}.

\section*{Discussion}

\noindent
Our experiments reveal that the main limiting factor of our CV-QKD protocol is the total noise, which is composed of the coupled noise and the amplification noise. Reduction of the latter, equivalent to a higher quantum efficiency of Bob's SQMs, results in a straightforward improvement of the protocol performance. Another route is to increase the codebook variance by adding trusted noise on Alice's side or by increasing the squeezing level. This process is limited by compression effects of our JPAs, which typically set on at input signal powers around $-130$~dBm. Travelling-wave parametric amplifiers~\cite{Macklin2015} could serve as alternative phase-sensitive amplifiers in future experiments, commonly tolerating higher input powers with quantum efficiencies comparable to our JPAs. Their broadband amplification properties enable the implementation of signal multiplexing techniques, which would also deliver significantly higher secure bit rates.

Our experiments show that SQMs implemented with phase-sensitive amplifiers can be considered as a microwave equivalent of optical homodyne detection. More precisely, our experiment demonstrates the possibility of using these SQMs to unravel properties of quantum states, particularly relevant for quantum state tomography ~\cite{Mallet2011,Knyazev2018}. This approach can be further extended to non-Gaussian state tomography and complements GKP error correction codes by offering a single-shot single-mode quadrature detection technique~\cite{Gottesman2001,hanamura2023}, necessary for these codes. Lastly, using the Shannon-Hartley theorem with the experimental parameters of the $2\textsuperscript{nd}$ run and our measurement bandwidth of $\SI{400}{\kilo\hertz}$, we estimate an upper bound of our experimental raw secret key rate up to $304~\text{kbit/s}$ for the lowest coupled noise $\bar{n}$, paving the way for secure high-bit-rate microwave CV-QKD communication. In particular, our demonstrated results promote the on-going development of local microwave networks~\cite{Awschalom2021,renger2023cryogenic}, where short-distance secure microwave quantum communication platforms could complement current classical microwave communication technologies such as Wifi and Bluetooth due to the intrinsic frequency and range compatibilities.
	
\section*{Methods}
\subsection*{Experimental squeezed microwave states}
\noindent
We experimentally generate squeezed states with JPAs, which are flux-tunable superconducting devices consisting of a harmonic $\lambda/4$ resonator shorted to ground with a dc-SQUID made of Al/AlO$_\mathrm{X}$/Al Josephson junctions. These JPAs are operated in the phase-sensitive regime pumping them with strong coherent microwave tones. The squeezed states are described using the squeeze operator $\hat{S} = \text{exp}((\xi^{\star}\hat{a}^2-\xi(\hat{a}^{\dagger})^2)/2) $, where $\hat{a} = \hat{q} + i\hat{p}$ ($\hat{a}^{\dagger} = \hat{q} - i\hat{p}$) is the annihilation (creation) operator with the quadrature operators $\hat{q}$ and $\hat{p}$ such that $\left[\hat{q}, i\hat{p} \right] = 1/2 $ and $\xi = r \text{e}^{i\varphi}$ is the complex squeezing amplitude. Here, the phase $\varphi = -2\gamma$ is related to the squeezing angle $\gamma$ between the anti-squeezed quadrature and the \textit{p} quadrature in the quadrature phase space. Additionally, $r$ represents the squeeze factor, related to the amount of squeezing. The latter is quantified using the squeezing level $S = -10 \log_{10}\left(\sigma_{\mathrm{s}}^2/0.25 \right) $. Similarly, we define the anti-squeezing level $A = 10 \log_{10}\left(\sigma_{\mathrm{as}}^2/0.25 \right) $. In our measurements, we implement a phase-locked loop with a feedback, which periodically adjusts the phase of our pump tones to maintain a stable squeezing angle ~\cite{Fedorov2018finite}.
\subsection*{Wigner tomography}
\noindent
Wigner function tomographies are performed using a reference state tomography based on measured quadrature moments associated with the to-be-reconstructed quantum state~\cite{Eichler2011, Menzel2012}.

\subsection*{Quadrature model and calibration measurements}	
\noindent
The microwave CV-QKD protocol is modelled by describing each element presented in the experimental schematic in Fig.\,\ref{fig:Fig_1} with a corresponding operator. The squeezing operation from the first JPA is described by a squeeze operator $\hat{S}_{\mathrm{A}}$. Each directional coupler is modelled with a beamsplitter operator, $\hat{C}_{\mathrm{A}}$ and $\hat{C}_{\mathrm{E}}$, and their associated power transmissivity, $\tau_{\mathrm{A}}$ and $\tau_{\mathrm{E}} = 1-\varepsilon_{\mathrm{E}}$, respectively. For the measurement JPA, we use a noisy squeeze operator, $\hat{S}^{'}_{\mathrm{B}}$, to account for the added noise $\bar{n}_{\mathrm{J}}$ of the JPA. Since we are considering single-shot measurements, we also include the HEMT amplifier, described by an amplification operator $\hat{H}$ to account for an amplification noise $\bar{n}_{\mathrm{H}}$. Additionally, we introduce path losses in between each component which are described by a beamsplitter operator $\hat{L}_i$ for $i \in \left\lbrace 1,2,3,4\right\rbrace$. The final output state after the HEMT can be expressed as 
\begin{equation}
	\label{eq:model_operator}
	\begin{gathered}
		\hat{\rho}_{\mathrm{out}} =  \hat{T} \hat{\rho}_{\mathrm{in}} \hat{T}^{\dagger} \text{,} \\
		\hat{T} = 	\hat{H} \hat{L}_4 \hat{S}^{'}_{\mathrm{B}}\hat{L}_3 \hat{C}_{\mathrm{E}} \hat{L}_2\hat{C}_{\mathrm{A}}\hat{L}_1\hat{S}_{\mathrm{A}} \text{,}
	\end{gathered} 
\end{equation} 
where $\hat{\rho}_{\mathrm{in}}$ is the overall input state of our experimental setup, accounting for signal modes and all other modes involved with the action of the operators.
All experimental parameters used in Eq.\,\ref{eq:model_operator} are extracted from independent calibration measurements, where we perform full Wigner tomography of the measured signals, under the assumption that all quantum states are Gaussian, to obtain the parameters individually (see Supplementary Information). The accuracy of our tomography method relies on a precise photon number calibration performed using Planck spectroscopy measurements~\cite{gandorfer2023twodimensional}.
\subsection*{Holevo quantity}
\noindent
The Holevo quantity of Eve, giving an upper bound on her accessible information about Alice's key, is computed as
\begin{equation}
	\chi_{\mathrm{E}} = S_{\mathrm{N}} \left(\int_{\mathrm{A}} d\alpha\, f\left(\alpha\right) \hat{\rho}_{\mathrm{E},\alpha} \right) - \int_{\mathrm{A}} d\alpha\, f\left(\alpha\right) S_{\mathrm{N}}\left( \hat{\rho}_{\mathrm{E},\alpha} \right) \text{,}
\end{equation}
by integrating over the ensemble of states which Eve obtains after her entangling cloner attack, described individually by a density matrix $\hat{\rho}_{\mathrm{E},\alpha}$. The function $f$ represents the probability density function of Alice's random variable. Here, the integral is taken over the ensemble of displacements that Alice can use during the communication and $S_{\mathrm{N}}$ is the von Neumann entropy.

\section*{Acknowledgements}
\noindent
We acknowledge the fruitful discussions and contributions of Philipp Krüger, Valentin Weidemann, and Sebastiano Covone.
\noindent

We acknowledge support by the German Research Foundation via Germany's Excellence Strategy (EXC-2111-390814868), the German Federal Ministry of Education and Research via the project QUARATE (Grant No.\ 13N15380), the project QuaMToMe (Grant No.\ $\text{16KISQ036}$), JSPS KAKENHI (Grant No.\ 22H04937) and JST ERATO (Grant No.\ JPMJER1601). This research is part of the Munich Quantum Valley, which is supported by the Bavarian state government with funds from the Hightech Agenda Bayern Plus.

\def\bibsection{\section*{\refname}} 
\bibliography{FF_arxiv_submission}
	
\end{document}